\documentclass[namedreferences]{solarphysics}
\usepackage[hyperref,optionalrh]{spr-sola-addons} % For Solar Physics
\usepackage{graphicx}                    % For eps figures, newer & more powerfull
\usepackage{multirow}
\usepackage{color}                       % For color text: \color command
\usepackage{breakurl}                         % For breaking URLs easily trough lines

\begin{document}

\begin{article}

\begin{opening}

\title{Analysing the effects of apodizing windows on local correlation tracking using
Nirvana simulations of convection}

\author{Rohan E. Louis$^{1}$\sep
        B. Ravindra$^{2}$\sep
        Manolis K. Georgoulis$^{3}$\sep
        Manfred K\"uker$^{1}$
       }
\runningauthor{Louis, Ravindra, Georgoulis \& K\"uker}
\runningtitle{The effects of apodizing windows on local correlation tracking}

\institute{$^{1}$ Leibniz-Institut f\"ur Astrophysik Potsdam (AIP),
	  An der Sternwarte 16, 14482 Potsdam, Germany
          email: \url{rlouis@aip.de}\\
          $^{1}$ Indian Institute of Astrophysics,
          Koramangala, Bengaluru 560034, India \\
	  $^{3}$ Research Center for Astronomy and Applied Mathematics of the Academy of Athens,
          4 Soranou Efesiou Street, Athens GR-11527, Greece\\
          }

\date{Received ... 2012 / Accepted ...}

\begin{abstract}
We employ different shapes of apodizing windows in the local
correlation tracking (LCT) routine
to retrieve horizontal velocities using numerical simulations of
convection. LCT was applied on a time sequence of temperature maps
generated by the Nirvana code with four different apodizing windows,
namely--Gaussian, Lorentzian, trapezoidal and triangular, with
varying widths. In terms of correlations (between the
LCT-retrieved and simulated flow field), the triangular and
the trapezoidal perform the best and worst, respectively.
On segregating the intrinsic velocities in the simulations on the
basis of their magnitudes, we find that for all
windows, a significantly higher correlation is obtained for the
intermediate and high-velocity bins and only modest or weak values in
the low-velocity bins. The differences between the LCT-retrieved
and simulated flow fields were determined spatially which show
large residuals at or close to the boundary of granules.
The extent to which the horizontal flow vectors retrieved by LCT
compare with the simulated values, depends entirely on the width of
the central peak of the apodizing window for a given $\sigma$.
Even though LCT suffers from a lack of spatial content as seen
in simulations, its simplicity and speed could serve as a viable
first-order tool to probe horizontal flows--one that is ideal for
large data sets.
\end{abstract}


\end{opening}

\section{Introduction}
\label{intro}
High-resolution observations of solar granulation from the
balloon-borne solar observatory \textit{SUNRISE}
\citep{2011SoPh..268....1B} reveal sub-structures such as
granular lanes which typically consist of a bright and a dark
edge, that travel from the visible boundary of granules into
the granule itself \citep{2010ApJ...723L.180S}. By comparing
the observations with numerical simulations,
\cite{2010ApJ...723L.180S} interpret these structures to
be signatures of vortex tubes which are a fundamental structure
element of turbulence and are important for transferring energy
from large to small scales. On the other hand, granules are
also considered good tracers of the large-scale flows
\citep{2014A&A...561L...6S}, particularly that of
supergranular flows \citep{1956MNRAS.116...38H}.
\cite{1962ApJ...135..474L} and \cite{1964ApJ...140.1120S}
showed that the edges of supergranular cells coincide with
the chromospheric emission network. This suggests that the
large-scale horizontal flow is responsible for advecting
magnetic fields along the network boundaries. Solar
granulation thus serves as a proxy for the transport of
small-scale magnetic fields and their subsequent
accumulation and intensification at the network. Hence it
is important to have tools that can identify and track
granular flows on the solar surface.

There are several methods to track solar granulation. Of
these, local correlation tracking
\citep[LCT;][]{1986ApOpt..25..392N,1988ApJ...333..427N,
2004ApJ...610.1148W,2008ASPC..383..373F} is the oldest
and most commonly used routine. LCT computes the relative
displacement of small sub-regions centred on a particular
pixel. A Gaussian window, whose full-width-at-half-maximum
(FWHM) needs to be roughly the size of the structures that are to
be tracked, is used to apodize the sub-regions. Thus, the
horizontal speed at each pixel can be determined knowing
the displacement, image scale, and the time interval. The
``balltracking'' method of \cite{2004A&A...424..253P}
also produces results with the same accuracy as LCT but
is significantly more efficient, computationally. On the
other hand, the feature-tracking routine of
\cite{1995ESASP.376b.213S} allows the determination
of physical characteristics of individual small features.
Other tracking routines include the induction method
\citep[IM;][]{2002ApJ...577..501K}, inductive local
correlation tracking \citep[ILCT;][]{2004ApJ...610.1148W},
Fourier local correlation tracking \cite[FLCT;][]{2004ApJ...610.1148W},
minimum-energy fit \citep[MEF;][]{2004ApJ...612.1181L},
minimum-structure reconstruction \citep[MSR;][]{2006ApJ...636..475G} method,
differential affine velocity estimator \citep[DAVE;][]{2006ApJ...646.1358S},
and nonlinear affine velocity estimator \citep[NAVE;][]{2005ApJ...632L..53S}.
All these techniques, except FLCT, are based on the ideal
induction equation and are used to track the
magnetic footpoint velocity using magnetograms.
\cite{2007ApJ...670.1434W} performed tests with the
above routines (except NAVE) on synthetic magnetograms
generated from anelastic MHD simulations of
\cite{1999ApJS..121..247L} and concluded that DAVE estimated
the magnitude and direction of velocities slightly more
accurately than the other methods, while MEF's estimates
of the fluxes of magnetic energy and helicity were far
more accurate than the others. While the above methods have
varying computational demands, LCT is the fastest routine
among these.

\cite{2013A&A...555A.136V} used simulated continuum images
from the CO$^5$BOLD code \citep{2012JCoPh.231..919F} to
compare plasma flows with those retrieved by LCT. A similar
analysis was performed by \cite{2014A&A...563A..93Y}
using synthetic continuum maps generated by the STAGGER
code \citep{2012A&A...539A.121B}. The above studies reveal that
the horizontal proper motions are not effectively retrieved
by LCT. This drawback has been attributed to the nature of
granules whose motions are representative of the large-scale
plasma flow for length and time scales greater than 2.5~Mm
and 30~min, respectively \citep{2001A&A...377L..14R}. While
the comparison of horizontal flows retrieved from LCT on
synthetic data is the best way to validate the method and
to determine its limitations, it is not overtly clear if the
proper motions derived from LCT can and should be compared
directly with simulated horizontal plasma motions. Firstly,
LCT tracks contrast fluctuations between two images which is
not the same as the underlying plasma motions.
Secondly, LCT only determines the relative displacement between
two images, that are separated in time and for a pixel of
interest. The resultant morphology would thus
reflect displacements and not velocity, despite using a
constant scaling factor (arising from the spatial sampling and
time step) for all pixels in the field of view.
Thirdly, LCT is formally not consistent with the normal
component of the induction equation, which can
be expressed as a continuity equation; but rather with the
advection equation \citep{2005ApJ...632L..53S}. Furthermore, it
is not known if a Gaussian apodizing window is or would be the best
choice when using LCT.

In this paper, we evaluate the
performance of LCT on convection simulations for different
shapes of apodizing windows, taking advantage of the
fact that LCT is the simplest and one of the most computationally
efficient velocity determination tools. Our aim is to
determine whether a different apodizing function can
significantly improve LCT results which could then be used
in future studies. Such a study, to the best of our
knowledge, has not been carried out previously. As a secondary
product of these tests, we also validate LCT and discuss its
advantages and limitations with respect to simulations, as well
as its implications for solar observations.

\section{Nirvana simulations of convection}
\label{nirv}
\begin{figure}
\centerline{
\includegraphics[angle=90,width = \columnwidth]{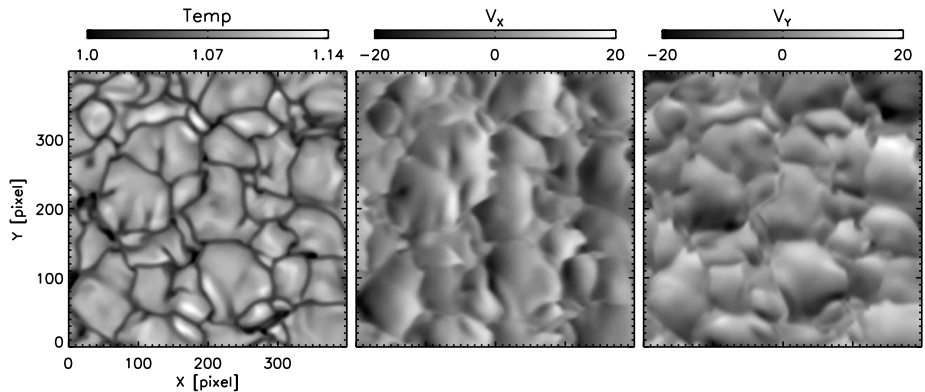}
}
\vspace{-10pt}
\caption{Left panel: Snapshot of temperature from the Nirvana
simulation. The corresponding {\em x-} and {\em y-}components
of the velocity vector are shown in the middle and right panels,
respectively.}
\label{fig01}
\end{figure}

The two-dimensional (2D) images showing temperature and horizontal
velocity components in Figure~\ref{fig01} are snapshots from a three-dimensional box simulation of magneto-convection in a
stratified gas layer. The simulation box is rectangular and
contains two layers of gas. The temperature stratification
is convectively unstable in the upper layer and stable in
the lower layer. The Rayleigh number in the unstable layer
is $10^7$, the thermal and magnetic Prandtl numbers both
assume values of 0.1. Density and temperature both assume
values of the order one while the isothermal sound speed
is 100 and the pressure increases from 10$^4$ at the top
to 10$^5$ at the bottom. The magnetic field is vertical and
the field strength has been chosen to be near equipartition
with the gas density at the top of the box.

The setup is the same as used in \cite{2012A&A...546A..23R}.
We use the Nirvana MHD code \citep{2004JCoPh.196..393Z}
to solve the equation of motion, the induction equation, and
the conservation equations for mass and energy in Cartesian
coordinates for an ideal gas of constant molecular weight.
The aspect ratio of the simulation box is $8 \times 8 \times 2$
in the $x$, $y$, and $z$ coordinates. Stratification is along
the $z$ axis with the gravity vector pointing in the negative
direction, \textit{i.e.} the $z$ axis pointing upwards. The bottom of
the box is at $z=-2$, the top at $z=0$. The horizontal cross
sections were taken at $z=-0.05$.

The boundary conditions are periodic in the horizontal
($x$ and $y$) directions. The vertical ($z$) boundary conditions
prevent outflow, \textit{i.e.} specify zero mass flow across the boundaries,
and stress-free in the horizontal components. The
temperature is kept fixed at the upper boundary while the heat
flux is prescribed at the lower boundary. We use a mesh with
$513 \times 513 \times 128$ points. The computations have been
carried out in the AIP's Leibniz cluster using the message
passing interface with 256 CPU cores in parallel. Dimensionless
units have been used for all quantities.

The time separation in the simulations was chosen to be
sufficiently smaller than the characteristic time scale of the
granules and to have a high correlation between successive
frames, so as perform tracking on consecutive
temperature maps.
The temperature distribution exhibits two clear peaks which
correspond to the dark cell boundaries and the granules which
represent upflowing plasma. Since temperature is a
dimensionless quantity, converting it into bolometric
intensity, to derive the image contrast, is not possible. A
proxy for the contrast was however done in the following manner.
The two-peak distribution was fitted with two Gaussians,
whose mean and $\sigma$ are represented as $m_1$, $\sigma_1$ and
$m_2$, $\sigma_2$, respectively. If $<I>$ represents the mean
value of the image then the contrast is expressed as
$((m_2 + \sigma_2) - (m_1 - \sigma_1))/<I>$. Using this
expression, the contrast is estimated to be about 6\%. In
comparison the contrast for solar granulation is about 14\% at
630 nm as shown by \cite{2008A&A...484L..17D}
using MHD simulations.

\begin{figure}
\centerline{
\includegraphics[angle=90,width = 0.8\columnwidth]{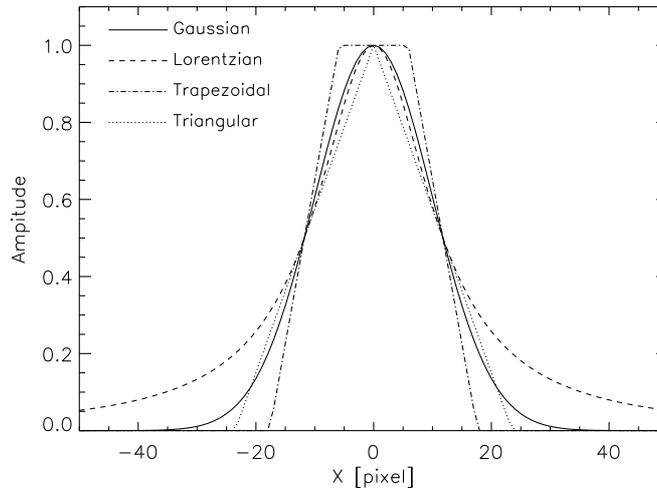}
}
\vspace{-10pt}
\caption{Different shapes of apodizing windows used in LCT.
The 1D profiles correspond to a $\sigma=10$ pixel.}
\label{fig02b}
\end{figure}

\section{Apodizing windows for LCT}
\label{apod}
Inferring the velocity of features from a set of two images
depends on two parameters, namely, the apodizing window
function and the time difference between the two images.
Before computing the cross correlation, the first step is
to generate the apodizing window function with a given
width ($\sigma$). The window function is multiplied
with sub-images in the reference frame and the object frame,
thus smoothing both the sub-images.
The apodization also restricts the motion of the features
to the window size. The choice of the time
step depends on the evolution time scale of the feature, as
a small time step might yield a poor correlation owing to a
small shift, while a large time step could render the
feature outside the apodizing window. Thus, a compromise between
the width of the apodizing window and the time
difference is necessary.

\begin{figure}
\centerline{
\includegraphics[angle=90,width = 1\columnwidth]{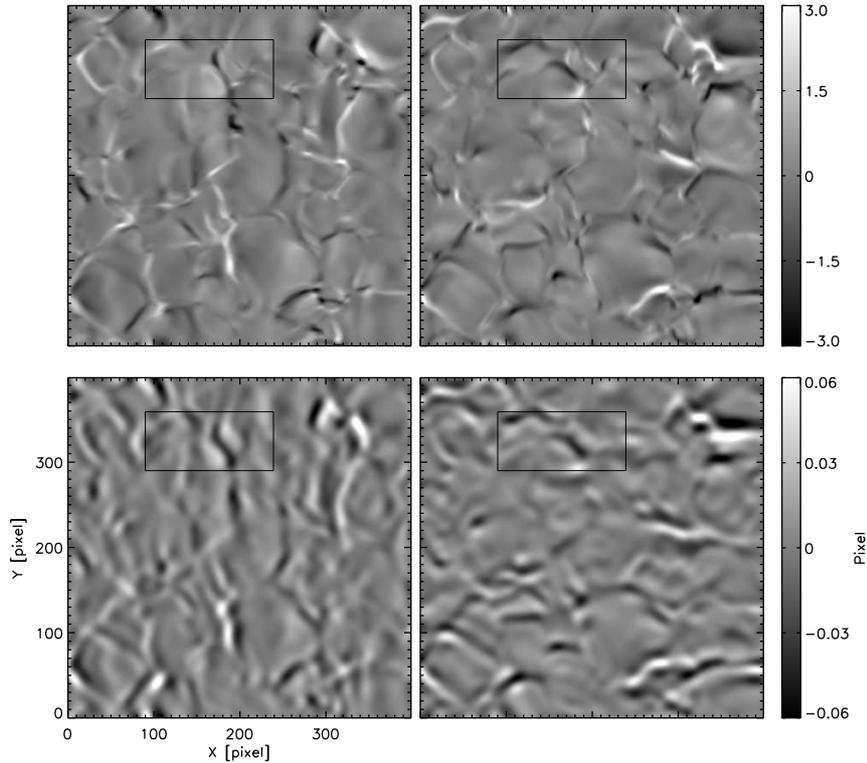}
}
\vspace{-10pt}
\caption{Top panels: Velocity-difference maps in {\em x-}(left)
and {\em y-}directions (right) obtained from successive frames
of the simulation sequence. Bottom panels: Example of horizontal
velocities obtained after LCT was applied to the corresponding
temperature maps. A Gaussian window of $\sigma=10$ pixel was
used in this case. The {\em black rectangle}
in the panels corresponds to a smaller region shown in
Figure~\ref{fig04}.}
\label{fig02}
\end{figure}

In order to compute the shifts, we used a Fast Fourier Transform
(FFT) based cross-correlation program \citep{2003ApJ...587..458N}.
The displacements were determined with sub-pixel accuracy by fitting
a quadratic function to the cross-correlated values near the peak.
This procedure yields displacement vectors for all pixels in each frame
and for the entire time sequence. The program is very fast and on
a 3.2 GHz Intel-core processor, it takes about 70~sec for a Gaussian
apodizing window having a width of $\sigma=15$ pixel and for an
image of size 513$\times$513 pixel. We generated 2D Gaussian,
Lorentzian, trapezoidal and triangular windows for widths of
$\sigma=$~10, 15, and 20 pixel. The windows were constructed
such that for a given $\sigma$, the FWHM
was the same for all, thus providing a basis for comparison. This
is shown in Figure~\ref{fig02b} for a one-dimensional (1D) plot of
each window function for a width of $\sigma=10$ pixel.

\begin{figure}
\centerline{
\includegraphics[angle=90,width = \columnwidth]{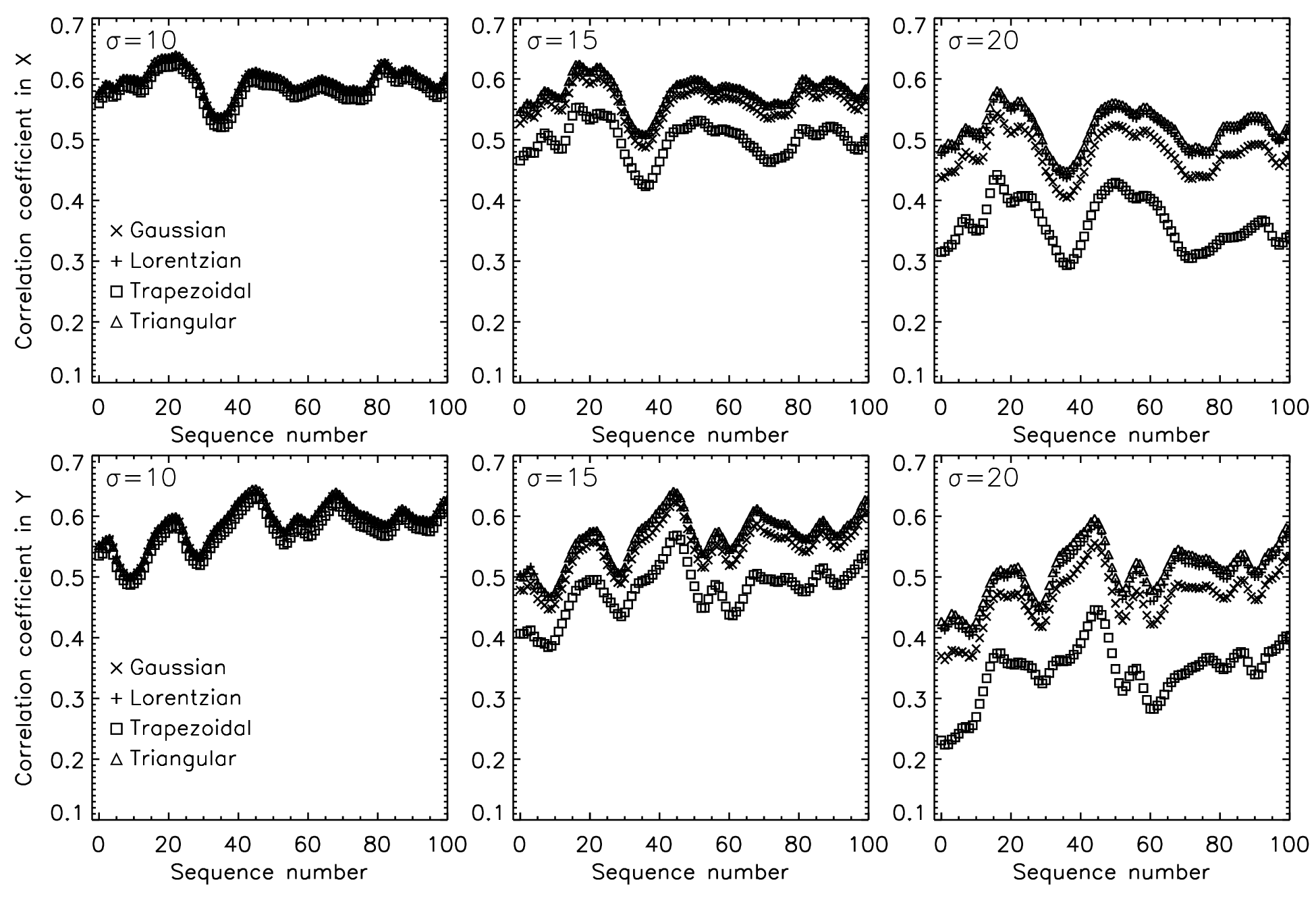}
}
\vspace{-5pt}
\caption{Linear correlations between simulated velocities and
that derived by LCT. Top panels: Correlation coefficients
in the $x$-direction as a function of time for different
apodizing windows, shown with different symbols and inscribed
in the left-most panel. Panels from left to right correspond
to different widths of the apodizing window. Bottom
panels: Same as above but for the $y$-direction.}
\label{fig03}
\end{figure}

As mentioned in Section~\ref{nirv} the tracking was done on successive
frames as we were interested in retrieving the instantaneous
horizontal motions. The central 400$\times$400 pixel region
was chosen for analysis to avoid the edges that were affected
by the apodization. The top panels of Figure~\ref{fig02} show
running-difference maps of the simulated horizontal velocity from
successive frames, while the bottom panel shows the displacements
determined by LCT using a Gaussian apodizing window with a width
of $\sigma=10$ pixel. The running-difference velocity maps, thus,
reflect displacements, albeit with time as a dimensionless scale
factor, and henceforth we will refer to them as velocities,
unless otherwise stated.

\section{Results}
\label{res}
\subsection{Correlations between LCT-retrieved and simulated horizontal velocities}
\label{compare}
We compute the Pearson linear correlation coefficient
(CC) of the horizontal velocities retrieved by LCT and
simulations for the different apodizing windows and
widths for the entire time sequence. Figure~\ref{fig03} shows
that the correlations in both $x$- and $y$-directions have a
time-dependent behaviour which is reflected very similarly
in all apodizing windows and for all widths considered.
This time-dependence of the CCs
is related to the temporal evolution of the
convective pattern seen in the simulated velocity-difference
maps. The correlation decreases as the
width of the apodizing window increases. For the
narrowest window having a width of $\sigma=10$~pixel,
the correlation is the highest and nearly the same
for all the apodizing windows. However, as the width of
the window increases, we find a distinction in the
correlations, the highest being for the Lorentzian
and triangular apodizing windows, followed closely
by the Gaussian and being the lowest for the
trapezoidal window.

The maximum CCs for the Gaussian, Lorentzian, and triangular
windows are about 0.64 in both the
$x$- and $y$-directions for a width of $\sigma=10$
pixel, while the corresponding minimum values are
0.52 and 0.50, respectively. Similar values are seen with
the trapezoidal window for the same width.
With the exception of the trapezoidal window, where the
minimum value of the CC is quite low for the largest
width, the same is relatively modest for the triangular,
Lorentzian and Gaussian windows. The difference between
the maximum (or minimum) CCs for $\sigma=10$ and 20~pixel
vary by about 20\% for the trapezoidal apodizing window,
while this variation is only about 10\% for the other
three windows. Although, we find a time dependence in
the CCs, the minimum and maximum values of the CC for the
triangular, Lorentzian and Gaussian apodizing windows
differ by about 15\%.

\begin{figure}
\centerline{
\includegraphics[angle=90,width = 1.15\columnwidth]{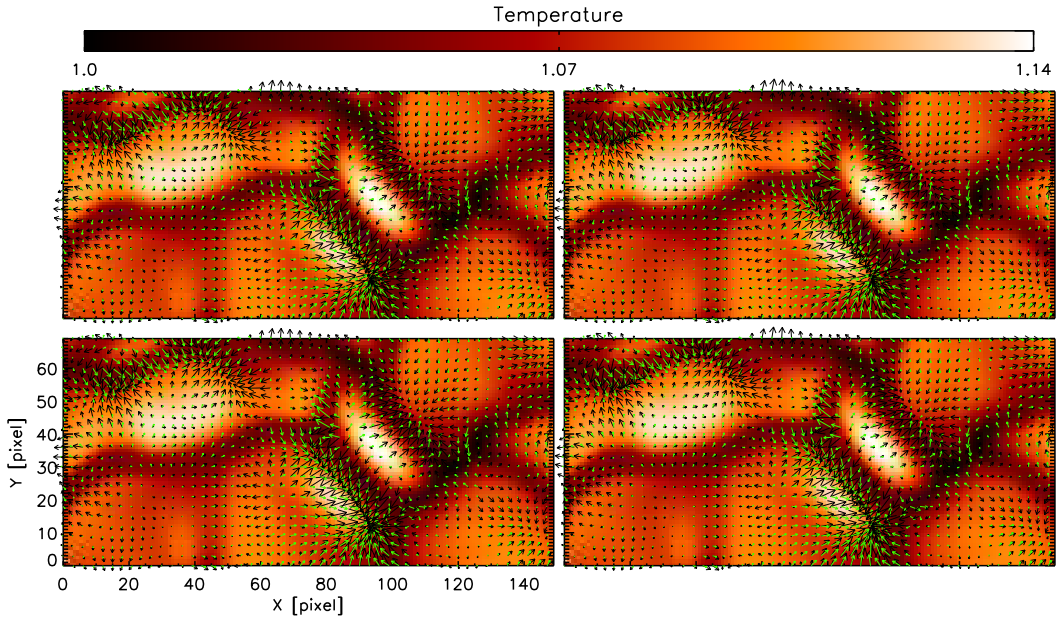}
}
\vspace{-15pt}
\caption{Comparison of horizontal flow vectors from LCT
({\em black arrows}) and simulations ({\em green arrows}) for
the area indicated by the {\em black box} in Figure~\ref{fig02}.
The arrows have been drawn for every third pixel and overlaid
on the corresponding temperature map. Top panels -- Gaussian (left)
and  Lorentzian (right)  apodizing windows. Bottom panels - trapezoidal
(left) and trianglular (right) apodizing windows.}
\label{fig04}
\end{figure}

\subsection{Comparison of flow vectors}
\label{scale}
As mentioned in Section~\ref{nirv}, the physical parameters
are dimensionless and in order to make a pixel-to-pixel
comparison of the simulated velocities with those from
LCT, we determined the scale factor relating the two
quantities, from a linear regression fit to the scatter
plot (not shown) in the $x$- and $y$-directions. The above
exercise was performed for each map in the time sequence.
Using the regression gradient as a scale factor, the
horizontal flow field from the simulations and that from
LCT can be directly compared and is illustrated in
Figure~\ref{fig04} for all windows with a width of $\sigma=10$
pixel. We find that the flow vectors from LCT are in concordance with
those from simulations. As seen earlier in Figure~\ref{fig02},
the horizontal velocities retrieved from LCT lack the fine
structure as observed in the simulated data although
there is an overall morphological agreement between the
two and this is the reason for the modest correlation
shown in Figure~\ref{fig03}. Subsequently, the LCT
flow vectors appear stronger than the simulated ones
particularly at locations where the intrinsic velocities
are high. The movie\footnote{Available as online material}
also shows the temporal evolution of the simulated flows,
which match the LCT-determined flows quite well. It is
observed that the flow maps retrieved by LCT using the
trapezoidal window are weaker in amplitude and spatially
smoother in comparison to the other three windows.

\begin{table}
\resizebox{0.95\textwidth}{!}{\begin{minipage}{\textwidth}
\caption{Correlation coefficients for different windows and
widths for different simulated velocity bins in the $x$- and
$y$-directions for the entire time sequence. Gau--Gaussian,
Lor--Lorentzian, Tap--trapezoidal, Tri--triangle.}
{\renewcommand{\arraystretch}{1.5}
\begin{tabular}{c|cccc|cccc|cccc}
\hline
\multirow{2}{*}{Velocity range}&\multicolumn{4}{|c|}{$\sigma = 10$} & \multicolumn{4}{|c|}{$\sigma = 15$}
&\multicolumn{4}{|c}{$\sigma = 20$} \\
\cline{2-13}
         & Gau  & Lor  & Tap  & Tri  & Gau  & Lor  & Tap  & Tri  & Gau  & Lor  & Tap  & Tri\\
\hline \hline
$|V_x|<0.5$               &  0.43 &  0.42 &  0.43 &  0.43 &  0.47 &  0.47 &  0.44 &  0.48 &  0.46 &  0.47 &  0.39 &  0.48\\
$0.5 \leq |V_x| \leq 1.5$ &  0.70 &  0.70 &  0.70 &  0.70 &  0.68 &  0.69 &  0.62 &  0.69 &  0.60 &  0.63 &  0.46 &  0.64\\
$|V_x|>1.5$               &  0.79 &  0.79 &  0.78 &  0.79 &  0.74 &  0.76 &  0.67 &  0.76 &  0.62 &  0.67 &  0.36 &  0.69\\
\hline
$|V_y|<0.5$               &  0.43 &  0.41 &  0.43 &  0.43 &  0.47 &  0.47 &  0.44 &  0.48 &  0.46 &  0.47 &  0.39 &  0.48\\
$0.5 \leq |V_y| \leq 1.5$ &  0.70 &  0.69 &  0.70 &  0.70 &  0.68 &  0.69 &  0.62 &  0.69 &  0.60 &  0.63 &  0.47 &  0.64\\
$|V_y|>1.5$               &  0.79 &  0.79 &  0.78 &  0.79 &  0.74 &  0.75 &  0.65 &  0.76 &  0.60 &  0.66 &  0.35 &  0.68\\
\hline
\end{tabular}
}
\label{tab2}
\end{minipage}}
\end{table}

\subsection{Velocity-dependent LCT values}
\label{velo}
Following Figure~\ref{fig04}, we investigate if the correlations
between simulated and LCT velocities depend on the intrinsic
magnitude of the flows since the values determined in
Section~\ref{compare} considers the entire distribution of
velocities. This test serves to check if LCT can recover
smaller velocities with the same accuracy as larger ones. From
the histogram of the simulated velocities, we selected three
bins corresponding to $|V|<0.5$, $0.5\le|V|\le1.5$ and $|V|>1.5$
which represent low, intermediate and high velocities, respectively
in both $x$- and $y$-directions. The three bins represent about
78.8\%, 19.2\% and 2.0\% of the 2D velocity distribution,
respectively in both directions and in time. Here, 2\% corresponds
to a sample of 3200 pixels. Note that the velocity bins described
above correspond to dimensionless values and were chosen by hand
from the velocity distribution. Table~\ref{tab2} summarizes the
CCs between the simulated velocities and those from LCT for the
three velocity bins and it is evident that when considered separately,
the correlations improve significantly, becoming strong especially
for the intermediate and high velocity bins. Thus, the modest
correlations obtained in Section~\ref{compare} are due to the low/weak velocities which dominate the distribution. The CCs for all windows
and for the smallest width of $\sigma=10$ pixel are about 0.70 and 0.79
for the intermediate and high velocity bins, respectively in
both $x$- and $y$-directions.
Even in the case of the largest width($\sigma=20$ pixel), the
CC, with the exception of the trapezoidal
window, is greater than 0.6 for both $x$- and $y$-directions.
The CCs are nearly independent of the width for all windows, in
the low velocity bin and in both directions, the variation
being less than 5\%.  Even though the spatial domain is
dominantly populated by weaker velocities, the global
regression coefficient, described in Section~\ref{scale},
is representative of the same from the intermediate velocity
bin.

\begin{figure}
\centerline{
\includegraphics[angle=90,width = 1.0\columnwidth]{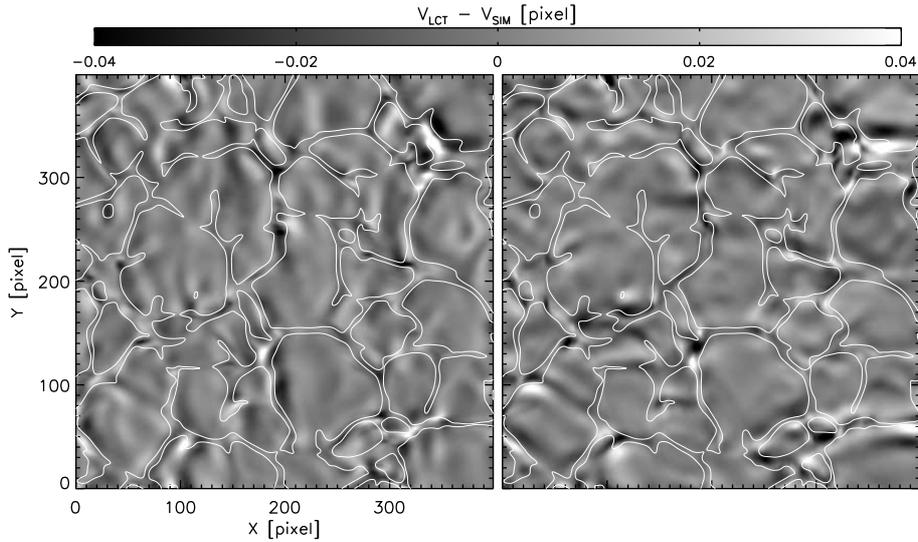}
}
\vspace{-20pt}
\caption{Left panel: Spatial difference maps between LCT
and simulated velocities in the $x$-direction for the
triangular apodizing window. Right panel: Same as
left panel but for the $y$-direction. The scaling of the
images is indicated in the color bar. The
maps correspond to a width of $\sigma=10$ pixel.
The contours correspond to a temperature of
1.05 and outline most of the intergranular lanes.}
\label{fig05}
\end{figure}

\subsection{Spatial distribution of velocity differences}
\label{spatial}
Figure~\ref{fig05} shows a difference map of the horizontal
velocities from simulations and those derived from LCT, in
$x$- and $y$-directions, using the triangular apodizing
window, for $\sigma=10$~pixel. This is also representative
of the maps obtained with the Lorentzian and Gaussian
windows of the same width. The relevant scale factor
described in Section~\ref{scale} was used for the simulated
velocity map and in the figure, black(white) represents
strong negative(positive) residuals. Using the contours from the
corresponding temperature map, we find that the locations
of good agreement are confined to granules which represent
upflowing material, while regions with strong residuals
are located at or close to the boundaries of the granular
cells. These residuals occur at specific locations in the
field of view and not at all cell boundaries. On a closer
inspection of the flow vectors obtained from LCT and those
from simulations reveal that there are subtle differences
between the two even at the location of granules but the
inherent, weak horizontal flows mask the small residuals
in comparison to that seen at the inter-granular lanes.

\section{Discussion}
\label{discuss}
We have used numerical simulations of convection to analyse
the performance of LCT for different shapes of apodizing
windows as such a study has not been carried out previously.
Since the horizontal velocities are known from simulations,
they can be compared with those retrieved by LCT and the latter
was applied to the simulated temperature maps so as to mimic
the situation often used in solar observations. We first
discuss the effects introduced by the different apodizing
windows and follow it up with the validation of LCT on
the simulated data.

\subsection{Performance of different apodizing windows}
\label{wind}
LCT was used on the simulations with four different shapes
of apodizing windows, namely--Gaussian, Lorentzian, trapezoidal,
and triangular, with different widths. The width of
the apodizing window is related to the size of the structure
to be tracked and hence will vary on the spatial resolution.
As the temperature maps from the simulations have significant
structural detail and are not affected by instrumental and
atmospheric effects, it is clear that the narrowest windows
(albeit still significantly larger than the size element)
would give better results. This was
indeed the case, as all four windows perform nearly the
same for the narrowest width, which is in agreement with
\cite{2013A&A...555A.136V}, who used a Gaussian
apodizing window function. Indeed, we have performed the above
tests for a width of $\sigma=5$~pixel and found that
although the results from all the windows yield the
same correlation coefficient, the values are less than
that observed in the $\sigma=10$ or 15~pixel case. The
lower correlations obtained with a width of $\sigma=5$~pixel,
imply that the displacement of the features exceeds
the width of the window. We also examined the case when
$\sigma$ was increased to 25~pixel and found that the
correlations decrease further than the $\sigma=20$~pixel
case with a similar separation of values amongst the four
windows. The above suggest that apodizing widths within
the range of 10 and 15 pixels are optimal for tracking
intensity features in the simulations.

The Gaussian window was only marginally inferior to the
triangular and Lorentzian windows in terms of the correlations
obtained. However the correlations from these three window functions
were less sensitive to the width than the trapezoidal window.
This could be attributed entirely to the width of the peak
centred on the pixel of interest, for a given $\sigma$. The
outer wings or edges of the window do not affect the results
as long as the window is wide enough to detect the
motion of the structure in the chosen time step.
The triangular window has an abrupt cut-off at its edge, while
its peak has the smallest width, and retrieves the same, albeit
marginally better, result than the Lorentzian and Gaussian windows,
both of which have comparatively extended wings. By the
same argument, the central peak or plateau of the trapezoidal
window increases more rapidly with $\sigma$, which would
explain why the correlations obtained with it are lower
than the others.

\subsection{Validation of LCT}
\label{valid}
The uniform image
quality and high spatial content in the simulations allowed us to
track consecutive frames thus eliminating the need to vary the time step.
In fact, the effectiveness of the
tracking routine is established by the accuracy of the
instantaneous flow field retrieved from successive frames
and has critical physical implications for understanding
energy transfer in the Sun at small, spatial, and temporal
scales \citep{2010ApJ...723L.180S,2011A&A...533A.126M}. As
the simulations presented here are dimensionless, it limits
a direct comparison in terms of physical units that can be
checked with those known from solar observations. This
limitation aside, the comparison of results from LCT and
simulations can be regarded as impartial and unbiased.

We reiterate that LCT only retrieves relative displacements
between two sub-images and as such should strictly be compared
with a running-difference of velocity and not directly with
velocity itself. Applying a scale factor to obtain velocity
from the LCT results will not in any way alter the nature of the
horizontal flows. This would explain why the results obtained
by \cite{2013A&A...555A.136V} vary significantly. Even though
the LCT flow maps appear to be a smoother version of that seen
in simulations, there is an overall morphological agreement,
which could be the reason for the moderate-to-high
correlations obtained in our analysis. Despite
this drawback, the fact that LCT is relatively simple and not
computationally intensive, makes it a viable diagnostic tool
for probing horizontal flows. The lack of fine structure
in the LCT horizontal flow maps can explain the results
of \cite{2014A&A...563A..93Y} who found that the Fourier
spectra for the LCT-determined
velocities is well below that from the actual velocity components.
We have not carried out any spatial smoothing of the simulated data
as \cite{2013A&A...555A.136V} but have performed a pixel-to-pixel
comparison. Our analysis also reveals that when the intrinsic
velocities are sufficiently strong, the accuracy of LCT is also higher
as reflected in the correlations corresponding to different velocity
bins.

The spatial differences in the LCT-determined flow fields
and simulations reveal that large residuals can occur
close to or at the edge of the granular boundaries with very
small differences at the granules themselves as also
seen in \cite{2007ApJ...670.1434W}. This suggests that
such sites could possibly reflect interesting
physical processes that could be the topic of further
investigation by employing other
computationally intensive tracking routines.

\section{Conclusions}
\label{conclu}
We have tested the conventional use of a Gaussian apodizing window
in LCT against other shapes of windows and found that for a given
$\sigma$, the width of the peak centred on the pixel of interest
influences the determination of the horizontal motions. On the
basis of this result, the triangular window yields the best result
while the trapezoidal, the worst. The LCT routine
was tested on simulations of convection but it remains to
be seen if a similar result can be obtained with magneto-convective simulations which replicate solar-like conditions
\citep{2005A&A...429..335V,2009Sci...325..171R,2010ApJ...720..233C}.
In the course of this analysis we also tested the validity of LCT
in determining known plasma motions and we conclude that LCT can
still be used as a first-order approach to derive horizontal
proper motions that is ideal for processing large data sets.

\begin{acks}
R. E. L is grateful for the financial assistance from the German
Science Foundation (DFG) under grant DE 787/3-1 and the European
Commission's FP7 Capacities Programme under Grant Agreement
number 312495. M. K. G acknowledges support by the European
Commission's FP7 Marie Curie Programme under grant agreement
no. PIRG07-GA-2010-268245. This work used the Nirvana code
developed by Dr. Udo Ziegler at the Leibniz-Institut
f\"ur Astrophysik Potsdam (AIP). We thank the referee for
the useful suggestions and comments.
\end{acks}

\bibliographystyle{spr-mp-sola}
\bibliography{louis_reference}

\end{article}

\end{document}